\documentstyle[12pt,aaspp4]{article}
\begin{document}
\parindent=1.0cm

\title {Near-Infrared Imaging of the Central Regions of Metal-Poor Inner 
Spheroid Globular Clusters}

\author{T. J. Davidge \altaffilmark{1}}

\affil{Canadian Gemini Office, Herzberg Institute of Astrophysics,\\
National Research Council of Canada, 5071 W. Saanich Road,\\ Victoria, 
British Columbia, Canada V9E 2E7 \\ {\it email:tim.davidge@nrc.ca}}

\altaffiltext{1}{Visiting Astronomer, Canada-France-Hawaii Telescope, which 
is operated by the National Research Council of Canada, the Centre National 
de la Recherche Scientifique de France, and the University of Hawaii}

\begin{abstract}

	$JHK$ images obtained with the Canada-France-Hawaii Telescope 
(CFHT) Adaptive Optics Bonnette (AOB) are used to investigate the near-infrared 
photometric properties of red giant branch (RGB) and horizontal branch (HB) 
stars in eight metal-poor globular clusters with R$_{GC} \leq 2$ kpc. The slope 
of the RGB on the $(K, J-K)$ CMDs confirms the metal-poor nature of 
these clusters, four of which (NGC 6287, NGC 6293, NGC 6333, and NGC 6355) 
have metallicities that are comparable to M92. The luminosity functions 
of RGB stars in inner spheroid and outer halo clusters have similar slopes, 
although there is a tendency for core-collapsed clusters to have slightly 
flatter luminosity functions than non-collapsed clusters. The distribution of 
red HB stars on the $(K, J-K)$ CMDs of inner spheroid clusters with [Fe/H] 
$\sim -1.5$ is very different from that of clusters with [Fe/H] $\sim -2.2$, 
suggesting that metallicity is the main parameter defining HB content among 
these objects. The RGB-bump is detected in four of the inner spheroid clusters, 
and this feature is used to compute distances to these objects. Finally, the 
specific frequency of globular clusters in the inner Galaxy is discussed in the 
context of the early evolution of the bulge. Based on the ratio of metal-poor 
to metal-rich clusters in the inner Galaxy it is suggested that the metal-poor 
clusters formed during an early intense burst of star formation. 
It is also demonstrated that if the globular cluster formation efficiency 
for the inner Galaxy is similar to that measured in other 
spheroidal systems, then the main body of the bulge could have formed 
from gas that was chemically enriched {\it in situ}; hence, 
material from a separate pre-enriched reservoir, such as 
the disk or outer halo, may not be required to form the bulge.

\end{abstract}


\section{INTRODUCTION}

	As the brightest members of the inner spheroid, globular clusters 
provide an important means of probing the structure and early evolution of the 
inner Galaxy. The most metal-rich inner spheroid clusters likely formed as part 
of the bulge (Minniti 1995a, C\^{o}t\'{e} 1999), and other sub-structures 
(Burkert \& Smith 1997), although some may belong to the classical halo 
(Heitsch \& Richtler 1999). Minniti (1996) argues that the metal-poor 
globular clusters in the central regions of the Galaxy are part of 
the classical halo, although the orbital properties of these objects 
are similar to those of bulge stars and gas in the inner Galaxy (C\^{o}t\'{e} 
1999). In addition, globular clusters in the inner Galaxy are more 
compact than those in the outer halo (van den Bergh 1994). It remains to be 
determined if the differences in orbital properties and sizes between 
metal-poor clusters in the inner spheroid and outer halo are primordial or the 
result of evolution.

	Efforts to study globular clusters in the inner spheroid 
are confronted with formidable observational challenges. 
Observations at visible wavelengths are complicated by the large, often 
non-uniform, extinction originating in the disk. The effects of 
this extinction can be reduced by observing at wavelengths longward of $1\mu$m, 
although then the temperature sensitivity of broad-band colors is decreased 
for all but the reddest stars. Contamination from disk and bulge stars 
poses a problem for studies at all wavelengths, and the failure to identify 
non-cluster stars can bias metallicity and distance estimates (Davidge 
2000a,b). Lacking spectroscopic information, contamination from non-cluster 
sources can be minimized by observing the dense central regions of 
clusters. While crowding then becomes an issue, stars at the main 
sequence turn-off (MSTO) can be resolved in the central regions of clusters 
at near-infrared wavelengths with 4 metre-class telescopes 
(Davidge \& Courteau 1999).

	Davidge (2000a) obtained broad and narrow-band near-infrared images of 
all known (as of 1996) globular clusters with [Fe/H] $\leq -1.3$ and R$_{GC} < 
3$ kpc, and used the properties of the brightest giants to compute 
the metallicities, reddenings, and distances of these objects. In the 
present study, $JHK$ images obtained with the CFHT AOB are used to investigate 
the photometric properties of stars as faint as the sub-giant branch (SGB)
in eight metal-poor inner spheroid clusters, selected to have [Fe/H] $< -1$ 
and R$_{GC} < 2$ kpc. The clusters are listed in Table 1, along with key 
properties taken from the June 1999 compilation of Harris (1996); 
in most cases the metallicities are from Zinn \& West (1984). The structural 
designations in the last column are from Trager, King, \& Djorgovski (1995). 
The properties of three outer halo clusters, spanning most of the 
metallicity range of the inner spheroid sample, and which are used as 
a control sample, are also listed in Table 1. The data for NGC 6287, M13, 
and M92 used in the current study were discussed previously by Davidge \& 
Courteau (1999). 

	Many of the inner spheroid clusters listed in Table 1 have been 
observed previously in the infrared. Minniti, Olszewski, \& Rieke (1995) 
presented $(K, J-K)$ color-magnitude diagrams (CMDs) of NGC 6287, NGC 6293, 
NGC 6333, NGC 6401, NGC 6522, and NGC 6642, while Davidge (2000a) presented 
$(K, J-K)$ CMDs and $K$ LFs covering the upper giant branch of all of the inner 
spheroid clusters in Table 1 except NGC 6401. The data discussed in the 
current paper go much fainter than these earlier studies; not only does this 
allow for a more detailed investigation of the properties of the RGB, which can 
be used to refine metallicity and distance estimates, but it also enables a 
preliminary reconnaisance of the cluster HBs.

	The paper is structured as follows. The observations and the data 
reduction procedures are described in \S 2, while the photometric measurements, 
CMDs, HB properties, giant branch LFs, and RGB-bump measurements are discussed 
in \S 3 -- 6. A brief discussion of the results follows in \S 7.

\section{OBSERVATIONS \& REDUCTIONS}

	The data were obtained with the CFHT AOB (Rigaut et al. 1998) and KIR 
camera during the nights of UT September 5, 6, and 7 1998. KIR 
contains a $1024 \times 1024$ Hg:Cd:Te array, with each pixel subtending 
0.034 arcsec on a side when used with the AOB.

	AO reference stars were selected at the telescope based on 
brightness, the absence of bright companions, and proximity to the 
cluster center. Once selected, the guide stars were 
centered on the detector to minimize the effects of anisoplanicity 
near the edge of the science field. In some cases this repositioning 
caused the cluster center to be offset to one quadrant of the detector.

	Each cluster was observed through $J, H,$ and $Ks$ filters. The 
telescope was offset between individual exposures in a $0.5 \times 0.5$ arcsec 
square dither pattern to facilitate the suppression 
of bad pixels and the construction of calibration 
frames from on-sky images. The main dataset for each cluster 
consists of exposures with 90 sec integration times recorded at each dither 
position. However, in many cases the brightest stars were saturated, so a 
second series of images with 3 -- 5 sec integration times were also recorded. 

	Standard stars from Casali \& Hawarden (1992) were also observed during 
the course of the three night observing run. The uncertainty in the 
photometric zeropoint of each filter derived from these observations is $\pm 
0.03$ mag.

	Deep exposures of NGC 6287 were obtained by Davidge \& Courteau (1999), 
and so only images with 5 sec integration times per dither position were 
recorded of this cluster during the September 1998 run. An AO 
reference star located 14 arcsec away from the reference star used by 
Davidge \& Courteau (1999) was selected so 
that the effects of anisoplanicity on the photometric measurements could 
be assessed (\S 3); details of the deep NGC 6287 observations, as well as 
the M13 and M92 observations that are also discussed in this paper, can be 
found in Davidge \& Courteau (1999).

	The data were reduced using the procedures described by Davidge \& 
Courteau (1999). The basic steps in the reduction process were: (1) subtraction 
of dark frames, (2) division by flat-field frames, and (3) the subtraction of 
thermal signatures and the DC sky level. The results were aligned to correct 
for the offsets introduced during acquisition, and then median-combined. 
The final $K$ images of the inner spheroid clusters observed during the 
September 1998 run are shown in Figures 1 and 2. In most cases the image 
quality is between 0.2 and 0.3 arcsec FWHM, and conspicuous 
signatures of anisoplanicity, such as the elongation of stellar images near 
the field edge (e.g. McClure et al. 1991), are not apparent.

\section{PHOTOMETRIC MEASUREMENTS}

	A single point-spread function (PSF) was constructed for each image 
using tasks in the DAOPHOT (Stetson 1987) photometry 
package, and stellar brightnesses were then measured 
with the PSF-fitting program ALLSTAR (Stetson \& Harris 1988). The finite size 
of atmospheric turbulence cells causes the PSF to vary with distance from the 
reference star (`anisoplanicity') on angular scales of a few arcsec, and so the 
use of a single PSF introduces position-dependent photometric errors. 
While it would be preferable to construct a variable PSF that tracks the 
affects of aniosplanicity, crowding and the complicated nature of the PSF 
variation with distance from the reference source make it difficult to do so.
In any event, previous studies of globular clusters with the AOB indicate that 
anisoplanicity does not introduce photometric uncertainties 
larger than a few percent over angular scales comparable to that of 
the KIR science field (Davidge \& Courteau 1999; Davidge 2000b).

	The effects of anisoplanicity can be investigated by comparing stellar 
brightnesses measured from images of a field recorded using different AO 
reference stars. The $K$ images of the March and September 1998 NGC 6287 
datasets in the vicinity of the reference stars are compared in Figure 3. 
The PSF changes most rapidly with location close to the reference stars,
and so a sample of moderately bright stars, which are marked in Figure 3, were 
selected to assess the impact of PSF variations on the photometry.

	Anisoplanicity will cause the difference in $K$ brightness, $\Delta K$, 
between the measurements from the March 1998 and September 1998 datasets to 
vary across the field. The two reference stars are separated by roughly 14 
arcsec, and in the top two panels of Figure 4 the histogram distribution of 
$\Delta K$ values for stars located within 7 arcsec of the September 
1998 reference star (interval 1) is compared with the distribution of stars 
located between 7 and 14 arcsec from this star (interval 2). The mean $\Delta 
K$ for all stars is $0.03 \pm 0.03$, confirming that the two datasets have 
consistent photometric calibrations. The short exposure time of the September 
1998 dataset introduces scatter that is significant for the faintest sources 
used in this comparison, which have $K = 14$; artificial star experiments (see 
below) indicate that the uncertainties introduced by crowding and noise at 
this brightness in the September 1998 dataset is $\pm 0.1$ mag. The relation 
between $K$ and $\Delta K$ is shown in the lower panel of Figure 4, and 
it is evident that the stars defining the two extremes in $\Delta K$ 
are among the faintest in the comparison. Photometric variability is 
another potential source of scatter.

	The mean values of $\Delta K$ in the 
two radial intervals are not significantly different: 
$\overline{\Delta K} = -0.02 \pm 0.04$ in interval 1, and $\overline{\Delta K} 
= 0.06 \pm 0.03$ in interval 2. While based on only a modest sample of stars, 
this comparison suggests that anisoplanicity introduces uncertainties 
of no more than a few percent in the $K$ photometric measurements of sources 
near the reference star, as originally suggested by Davidge \& Courteau (1999). 

	Artificial star experiments were used to assess sample completeness 
and estimate the effects of crowding and random noise on the photometry. 
Artificial stars were added to each final image using the ADDSTAR routine in 
DAOPHOT, which scales the PSF constructed for each dataset and adds 
random noise. The number of added stars was restricted to a 
small fraction of the number of stars actually detected to 
prevent artificially increasing the amount of crowding. Because a fixed PSF is 
assumed, these experiments do not monitor uncertainties introduced by 
anisoplanicity; however, the scatter in the CMDs is consistent with that 
predicted from the artificial star experiments (\S 4), indicating that 
anisoplanicity is not the dominant source of noise. 

	The artificial star experiments indicate that the long exposure $K$ 
data are typically complete to $K = 14$, with 50\% completeness occuring at 
$K = 17$. The photometric uncertainties are typically $\pm 0.02 - 0.04$ mag at 
$K = 14$, and $\pm 0.2 - 0.3$ mag at $K = 17$.

\section{COLOR-MAGNITUDE DIAGRAMS}

	The $(K, H-K)$ and $(K, J-K)$ CMDs of the inner spheroid clusters 
are shown in Figures 5 and 6, while the corresponding CMDs for the outer 
halo clusters are plotted in Figures 7 and 8. The photometry for 
stars brighter than $K = 12$ was obtained from the short exposure images. 
Images with short exposure times were not recorded for M92, and so the bright 
cut-off for this cluster is defined by detector saturation. 

	The main sequence turn-off (MSTO) for old metal-poor systems 
occurs near M$_K = 3$ (Bertelli et al. 1994), which corresponds to $K = 17.5 
- 18$ at the distance of the Galactic Center (Reid 1993; Davidge 2000a). 
Hence, the CMDs of the inner spheroid clusters, which typically extend to $K = 
18$, sample the entire giant branch, including the SGB, although incompleteness 
becomes significant at this point (\S 5). The CMDs for NGC 6287 are deeper 
than those of the other clusters because the data were recorded on a night 
with significantly better image quality. 

	The $H-K$ color is insensitive to effective temperature, and 
so the width of the $(K, H-K)$ CMD provides an independent means of assessing 
star-to-star photometric scatter. The standard deviation about the mean $H-K$ 
value in Figure 5 for stars with $K$ between 12 and 14 is typically $\pm 0.02$ 
to $\pm 0.03$ magnitude, suggesting that the random uncertainty in individual 
photometric measurements is on the order of $\pm 0.02$ magnitudes. This is 
consistent with the uncertainties predicted by the artificial star experiments, 
which do not account for PSF variations across the field, thus suggesting that 
anisoplanicity is not the dominant source of photometric scatter.

	Kuchinsky et al. (1995) investigated 
the relation between metallicity and the slope of the giant branch on $(K, 
J-K)$ CMDs, and the calibration of this relation has recently been extended to 
very metal-poor clusters by Ferraro et al. (2000) and Ivanov et al. (2000). The 
slope of the upper giant branch was measured from the $(K, J-K)$ CMDs in 
Figures 6 and 8 using the procedures described by Kuchinsky et al. (1995), 
and the results are listed in the second column of Table 2.

	The giant branch slopes of NGC 6287, NGC 6293, NGC 6333, and NGC 6355 
are similar to M92, confirming that these clusters are very 
metal-poor. The metallicities derived from the relation 
between [M/H] and giant branch slope specified in Figure 12 of Ferraro et al. 
(2000) are listed in the third column of Table 2, and for most of the clusters 
there is general agreement with the values listed in Table 1. Only a few stars 
define the giant branch sequence of NGC 288, and the central regions of this 
cluster are also devoid of the brightest giants that have a significant 
impact on slope measurements (Davidge 2000a); consequently, the RGB slope 
measured for NGC 288 from the current data is not well defined, and so a 
metallicity was not calculated.

	The HB and RGB sequences tend to be well separated on the $(K, J-K)$ 
CMDs. Only a modest number of HB stars are detected in each of the outer 
halo clusters, while significantly richer HB populations are found in the 
inner spheroid clusters; such a difference in stellar density is not unexpected 
given that inner spheroid clusters are more compact than outer halo clusters 
(van den Bergh 1994). The current data can be used to probe the HB content 
of the inner spheroid clusters, although the scatter in the CMDs becomes 
significant when $K = 17$, and this hinders efforts to study the bluest HB 
stars; for example, the extremely blue HB stars in NGC 6273 studied by Piotto 
et al. (1999) will have $K = 20 - 21$.

	Searle \& Zinn (1978) compared the HB morphologies of 
globular clusters in different regions of the Galaxy, and concluded that 
while the HB morphology of clusters in the inner Galaxy is defined by 
metallicity, this is not the case for clusters in the outer halo; more 
recently, the apparent absence of a second parameter affect among clusters 
in the inner Galaxy has been supported by Lee, Demarque, \& Zinn (1994). 
Given that age is one of the many (e.g. Salaris \& Weiss 1997, 
Buonanno et al. 1997) parameters that can affect HB content, 
then one possible explanation of these results is that the inner Galaxy formed 
over a shorter period of time than the outer halo, as might be 
expected if the Galaxy formed from the inside out (e.g. Larson 1990).
However, recent studies of globular clusters at very low Galactic latitudes 
challenge the long-held notion that HB morphology depends solely on 
metallicity within a few kpc of the Galactic Center (Davidge \& Courteau 1999, 
Rich et al. 1998; Ortolani et al. 1999).

	In an effort to investigate qualitatively the effect of 
metallicity on HB morphology among metal-poor inner spheroid clusters, 
composite CMDs were constructed by grouping the clusters according 
to metallicity, and then combining the CMDs of the clusters in each group. 
Two groups were considered: moderately metal-poor (NGC 6273, NGC 6401, NGC 
6522, and NGC 6642), and very metal-poor (NGC 6287, NGC 6293, NGC 6333, 
and NGC 6355). The HBs of NGC 6273 and NGC 6522 in Figure 6 are not 
well-defined, and so a composite CMD for moderately metal-poor 
clusters that did not include these objects was also constructed. 

	The CMDs in each group were corrected for cluster-to-cluster 
differences in reddening by registering the CMDs along the 
horizontal axis using the mean giant branch color at $K = 14$ 
as a reference. A corresponding shift to the vertical 
axis based on the Rieke \& Lebofsky (1985) extinction curve was also applied.
The inner spheroid clusters are roughly equidistant, so no effort was made 
to adjust for differences in distance.

	The composite CMDs are compared in Figure 9, and the HB in the [Fe/H] 
$= -2$ CMD is clearly different from the [Fe/H] $= -1.5$ CMDs computed 
with or without NGC 6273 and NGC 6522. In particular, the HB of the moderately 
metal-poor group consists of stars that are more or less uniformly distributed 
to the left of the giant branch with $\Delta K$ between 2 and 0. For 
comparison, the HB of the very metal-poor group is dominated by a clump of 
stars with a J-K color that is roughly 0.5 mag bluer than the giant branch. 
This simple comparison indicates that metallicity is the main parameter 
defining red HB content among metal-poor inner spheroid clusters, although the 
possibility that other parameters affect HB content of course can 
not be ruled out.

\section{GIANT BRANCH LUMINOSITY FUNCTIONS}

	The $K$ LFs of RGB stars, which were constructed from the CMDs after 
removing HB stars, are plotted in Figures 10 (inner 
spheroid clusters) and 11 (outer halo clusters). 
The LFs follow power-laws, and so can be characterized by an 
exponent, $x$, which was measured for each cluster by using the method of 
least squares to fit a power-law to the completeness-corrected data between $K 
= 12$ and 16. $K = 12$ was selected as the bright limit to avoid small number 
statistics near the RGB-tip, while $K = 16$ was adopted as the 
faint limit to avoid the onset of the sub-giant branch (SGB), which 
the models of Bertelli et al. (1994) predict should occur between $K = 16$ and 
17 in old stellar systems at the distance of the Galactic Center (GC).

	The $x$ values for the various clusters are listed in the second column 
of Table 3. Davidge (2000a) measured LF exponents for many of these clusters 
using images that, while covering a much larger area, sampled only the 
brightest cluster members, and his $x$ values are listed in the last column 
of Table 3. The mean difference between the two sets of exponents in Table 3 
is $\Delta x = 0.05 \pm 0.04$, indicating general agreement. 
However, significant differences occur for some clusters. 
For example, in the case of NGC 6287, which Davidge (2000a) noted would 
have a markedly different exponent if stars fainter than $K = 12.5$ were 
included, $\Delta x = -0.16 \pm 0.04$.

	The unweighted mean exponent for the eight inner spheroid clusters is 
$\overline{x} = 0.31 \pm 0.02$, which is not significantly different from the 
mean of the three outer halo clusters, which is $\overline{x} = 0.29 \pm 0.03$. 
Therefore, metal-poor clusters in the inner spheroid and 
outer halo have, on average, similar LF exponents. Nevertheless, 
the uncertainties in $x$ are typically $\pm 0.02 - 0.04$, and the entries 
for the inner spheroid clusters in Table 3 span a much larger range than this, 
suggesting that real cluster-to-cluster scatter might be present. 
Three of the inner spheroid clusters are core-collapsed, and 
the central regions of core-collapsed clusters tend to be devoid of 
bright giants (e.g. Djorgovski et al. 1991; Janes \& Heasley 1991; Djorgovski 
\& Piotto 1993; Davidge 1995, Burgarella \& Buat 1996). The absence of 
bright stars might be expected to affect the LF exponents, and hence 
introduce cluster-to-cluster scatter in the $x$ measurements -- is evidence for 
this seen in the current data?

	There is a remarkable degree of uniformity among the $x$ values for the 
core-collapsed clusters in this sample, with $\overline{x} = 0.27 \pm 0.01$ for 
NGC 6293, NGC 6355, and NGC 6522. For comparison, the unweighted mean of the 
remaining five inner spheroid clusters is $\overline{x} = 0.34 \pm 0.03$, 
while $\overline{x} = 0.32 \pm 0.02$ if the three outer halo clusters are also 
considered. These data hint that there may be a slight difference 
between the LF exponents of dynamically evolved and unevolved clusters, 
although there is clearly a need to sample a larger number of objects to 
confirm these results. The sense of the difference in $x$ between 
the dynamically evolved and unevolved clusters is somewhat surprising, as the 
absence of bright giants would be expected to produce a steeper LF, 
which is contrary to what is seen here. However, Davidge (1991) 
compared the optical and near-infrared color profiles of the globular cluster 
NGC 4147, which Djorgovski \& King (1986) suggest might be core-collapsed, and 
concluded that the giant branch population near the cluster center is depleted 
at brightnesses well below the RGB-tip, extending at least as faint as the HB; 
consequently, the depletion of stars near the RGB-tip might only be the most 
conspicuous signature of a process that occurs over a larger range of 
brightnesses. 

\section{THE RGB-BUMP}

	As a star ascends the giant branch the hydrogen-burning shell 
encounters a discontinuity in the abundance profile that marks the maximum 
inward extent of envelope convection (e.g. Iben 1968). The pace of evolution 
slows, with the result that a bump occurs on the differential LF. As 
metallicity drops, the so-called RGB-bump occurs at progressively more evolved 
states (e.g. Fusi Pecci et al. 1990, Ferraro et al. 2000), and hence becomes 
more difficult to detect due to the increased statistical flucuations in star 
counts. Deep mixing in stars on the lower giant branch can affect the amplitude 
of the RGB-bump (Langer, Bolte, \& Sandquist 2000). 

	The integrated LF is a powerful tool for detecting the RGB-bump (e.g. 
Lee \& Carney 1999). A least squares fit was made to the integrated $K$ LF of 
each cluster between $K = 13.5$ and $K = 16$, and the RGB-bump was identified 
as the point after which the LF departs from this relation at the bright end. 
The RGB-bump was detected in four inner spheroid clusters using this method: 
NGC 6273 ($K = 13.2 \pm 0.1$), NGC 6401 ($K = 13.2 \pm 0.1$), NGC 
6522 ($K = 13.4 \pm 0.1$), and NGC 6642 ($K = 13.4 \pm 0.1$), and the 
integrated LFs of these clusters are shown in Figure 12; the RGB-bump was not 
detected in the outer halo or the remaining inner spheroid clusters.

	The RGB-bump will also produce a feature in differential LFs, 
although the signature is more difficult to detect given the lower 
signal-to-noise ratio. To check the RGB-bump brightnesses measured above, the 
differential LF for each cluster was compared with the power-laws that were 
fit to the entire RGB (\S 5). Significant departures from 
the power-law relations were detected in the differential $K$ LFs 
of NGC 6273, NGC 6401, NGC 6522, and NGC 6642 at the same brightnesses as found 
in Figure 12. The differential LFs of these clusters are shown in Figure 13, 
with the RGB-bump marked. The comparison in Figure 13 verifies the RGB-bump 
brightnesses measured from the integrated LFs.

	The distance moduli of NGC 6273, NGC 6401, NGC 6522, and NGC 6642 
were computed using the RGB-bump calibration given in the top panel of Figure 
13 of Ferraro et al. (2000), and the results are listed in the second column of 
Table 4, where corrections for extinction have been made using the Rieke \& 
Lebofsky (1985) reddening law. Piotto et al. (1999) measured the $V$ brightness 
of the RGB-bump in NGC 6273, and found that $\mu_0 = 14.8$, which is in 
excellent agreement with the near-infrared RGB-bump distance modulus computed 
here. 

	Davidge (2000a) calculated distance moduli for NGC 6273, NGC 6522, and 
NGC 6642 using the brightnesses of the RGB-tip and the HB, and the results are 
listed in the third and fourth columns of Table 4. The mean difference between 
the distance moduli computed using the RGB-bump and RGB-tip is $0.3 \pm 0.4$, 
while the mean difference between the RGB-bump and HB results is $0.7 \pm 0.2$; 
in both cases the quoted uncertainties are the errors in the mean. The RGB-tip 
and HB distances given by Davidge (2000a) are based on the empirical Carney, 
Storm, \& Jones (1992) RR Lyrae calibration, while Ferraro et al. (2000) 
calibrate the RGB-bump using the theoretical ZAHB calibration given by Ferraro 
et al. (1999), which is consistent with {\it Hipparcos} sub-dwarf 
brightnesses. The Carney et al. (1992) and Ferraro et al. (1999) calibrations 
differ by roughly 0.2 mag at the metal-poor end, in the sense that the former 
predicts that RR Lyraes are intrinsically fainter. The application of a 0.2 mag 
correction to the RGB-tip and HB distance moduli in Table 4 brings these values 
into better agreement with the RGB-bump distances, although the spread between 
the various distance estimates for NGC 6642 is still very large.

	Of the three features used to measure distances in Table 4, the RGB-tip 
is likely the most prone to systematic errors, due to (1) the rapid pace of 
evolution near the RGB-tip and (2) the tendency for bright giants to be 
depleted near the center of dynamically evolved clusters. Both of these sources 
of error will bias the measured RGB-tip brightness, causing it to 
be fainter than the actual value, with the result that distances will be 
overestimated. In fact, the RGB-tip and HB distances given by Davidge (2000a) 
for 19 metal-poor inner spheroid clusters differ by $\Delta \mu = 0.43 \pm 
0.13$ mag, in the sense that the RGB-tip gives larger distances. The clusters 
that Trager et al. (1995) find are core-collapsed or have uncertain dynamical 
states tend to have the largest values of $\Delta \mu$, with an average 
$\Delta \mu = 0.56 \pm 0.20$. For comparison, the clusters that are not core 
collapsed have an average $\Delta \mu = 0.26 \pm 0.13$. Thus, this comparison 
suggests that dynamical evolution is the dominant source of systematic 
errors in the use of the RGB-tip as a distance indicator.

	Distances to low Galactic latitude clusters computed from the HB are 
also subject to systematic effects introduced by contamination from the field 
HB component, which will bias distances towards that of the peak in the bulge 
stellar density distribution along the sight line. The brightness of the HB 
also varies by a few tenths of a magnitude among clusters of the same 
metallicity, likely due to age (e.g. Lee \& Carney 1999). The RGB-bump offers 
a means of measuring distances to metal-poor bulge clusters that 
should be less susceptible to systematic errors. 
In particular, the RGB-bump occurs in a portion of the 
CMD where bulge giants can be identified using CO indices (Davidge 2000a), so 
that field stars can be removed from heavily contaminated cluster fields. 
Models predict that the brightness of the RGB-bump is only slightly sensitive 
to age, with $\frac{\Delta M_V}{\Delta t_{Gyr}} = 0.025$ mag Gyr$^{-1}$ 
(Ferraro et al. 1999). However, the RGB-bump is only 
detected in the most metal-rich clusters in the current sample, 
and data covering a significantly larger area will be required to detect the 
RGB-bump in the remaining, more metal-poor, clusters.

\section{DISCUSSION}

	$J, H,$ and $K$ images obtained with the CFHT AOB have been used to 
study the near-infrared photometric properties of evolved stars in eight 
metal-poor inner spheroid clusters. The CMDs show well-defined sequences, and 
artificial star experiments indicate that the scatter in 
these data is dominated by random errors. These data thus provide further 
evidence that photometric measurements with uncertainties of a few percent can 
be obtained of stars within 10 - 15 arcsec of the guide source in 
AO-compensated images. In \S 7.1 the results from the current work are 
discussed on a cluster-by-cluster basis, while in 
\S 7.2 the relative numbers of metal-poor and 
metal-rich clusters are used to probe the early evolution of the inner spheroid.

\subsection{Comparisons with Previous Work}

\vspace{0.3cm}

	Piotto et al. (1999) use $B$ and $V$ WFPC2 images to investigate the 
stellar content of NGC 6273, and point out that this cluster is in some 
respects similar to $\omega$Cen. Based on the width of the $(V, B-V)$ CMD, 
Piotto et al. suggest that an abundance dispersion may be present, although 
differential reddening on small angular scales likely also contributes 
to smearing of the giant branch. The width of the NGC 6273 giant branch 
on the $(K, J-K)$ CMD in Figure 6 is similar to that of the other 
clusters, while the star-to-star distribution of CO indices is 
not markedly different from that of other metal-poor clusters (Davidge 2000a). 
Consequently, near-infrared observations do not support the presence of a 
significant metallicity dispersion near the center of NGC 6273, suggesting that 
the smearing of the giant branch at visible wavelengths is a consequence of 
differential reddening.

	Rutledge, Hesser, \& Stetson (1997) measured the strength of the 
near-infrared Ca triplet in the integrated spectrum of NGC 6273 
and concluded that [Fe/H] $=-1.80 \pm 0.03$ using the 
Zinn \& West (1984) calibration, and $-1.53 \pm 0.05$ using the Carretta \& 
Gratton (1997) calibration. The metallicity of NGC 6273 derived from the 
slope of the giant branch with the current data is consistent with the latter 
result.

	As the most metal-poor cluster yet discovered in the inner 
spheroid, NGC 6287 has the potential of providing important clues 
into the early evolution of the inner Galaxy. 
Stetson \& West (1994) compared the $(V, B-V)$ CMD of NGC 6287 with 
that of the metal-poor halo cluster M15, and found that 
the HB brightnesses of these clusters differ by a few tenths of a mag, 
in the sense that NGC 6287 is the fainter of the two. 
The giant branch of NGC 6287 is steeper than that of other metal-poor 
clusters, such as M92, on the $(K, J-K)$ CMD (Davidge 1998; Davidge \& 
Courteau 1999); in fact, the CMDs in Figures 6 and 8 show a tendency 
for cluster giant branches to become shallower when $K < 12$, but for 
NGC 6287 the giant branch slope appears not to change near the RGB-tip.

	Given the extreme metal-poor nature of NGC 6287, it is important to 
determine if this cluster formed as part of the inner spheroid. The compact 
nature of NGC 6287 indicates that it is likely not an interloper from the 
outer halo (van den Bergh, Morbey, \& Pazder 1991). However, 
compactness is not an unambigous criterion for membership in 
the inner spheroid, since the globular clusters in the Fornax dwarf spheroidal 
galaxy are also compact (van den Bergh 1994). Consequently, 
the possibility that NGC 6287 may not have formed as part of the inner spheroid 
cluster system, but may have been stripped from a dwarf galaxy that was 
accreted by the Milky-Way (e.g. van den Bergh 2000), can not at this time be 
ruled out. 

	Janes \& Heasley (1991) used CCD data to investigate the $(V, B-V)$ 
CMDs of NGC 6293 and NGC 6333, and found that the giant branches of both 
clusters are well matched by that of M92, although Ferraro et al. (1999) note 
that the RGB locus of NGC 6333 defined by the Janes \& 
Heasley (1991) data crosses the giant branches of other clusters, 
possibly indicating uncertainties in the photometric calibration. 
The slopes of the giant branches of NGC 6293 and NGC 6333 on the $(K, J-K)$ 
CMDs are consistent with a very low metallicity. Janes \& Heasley 
(1991) also found that the upper giant branch of NGC 6293 is poorly populated, 
possibly due to dynamical evolution. The power-law exponent of the RGB $K$ LF 
of NGC 6293 is similar to that of the other core-collapsed clusters in this 
sample, which as a group tend to have flatter LFs than non-collapsed clusters.

	NGC 6355 is the least studied cluster in the current sample, and the 
only published CMD is that presented by Davidge (2000a), who found that [M/H] = 
--1.6 based on comparisons with theoretical isochrones. The slope of the RGB 
on the $(K, J-K)$ CMD in Figure 6 confirms that NGC 6355 is very metal-poor, 
and hence is a potentially important target for studies of the age of the inner 
spheroid.

	Barbuy et al. (1999) obtained $V$ and $I$ images of NGC 6401, and their 
CMD of stars between 1.3 and 2.0 arcmin from the cluster center shows a red HB 
and a curved upper giant branch similar to that in 47 Tuc. However, the slope 
of the giant branch on the near-infrared CMD published by Minniti et al. (1995) 
is comparable to that of M92, suggesting that NGC 6401 is very metal-poor. 
Minniti (1995b) obtained visible spectra of bright giants in NGC 6401, which 
radial velocity measurements suggest are cluster members, and concluded that 
[Fe/H] $= -1.1 \pm 0.2$. The slope of the NGC 6401 RGB on the $(K, J-K)$ CMD 
in Figure 6 indicates that [Fe/H] $= -1.5$. The current data also indicate 
that the HB of this cluster is uniformly populated over a range of $J-K$ 
colors, as expected if the cluster is old and moderately metal-poor. Thus, it 
appears that NGC 6401 is metal-poor, and not metal-rich.

	Because it is located in BW, NGC 6522 is the most extensively studied 
cluster in the inner spheroid. The CMD of NGC 6522 at visible wavelengths shows 
a giant branch with a color and slope characteristic of low metallicities 
(Terndrup \& Walker 1994; Barbuy, Ortolani, \& Bica 1994). Rutledge et 
al. (1997) use the depth of Ca II absorption in the intregrated cluster 
spectrum to assign [Fe/H] $= -1.50 \pm 0.05$ on the Zinn \& West (1984) 
metallicity scale, and [Fe/H] $= -1.21 \pm 0.04$ on the Carretta \& Gratton 
(1997) scale. The metallicity derived from the giant branch slope in the 
present study is consistent with the former measurement.

	Shara et al. (1998) used WFPC2 data to examine the $(B, U-B)$ CMD of 
stars near the center of NGC 6522, and found that the upper giant 
branch population is depleted within 9 arcsec of the cluster center. 
Rich et al. (1998) suggest that NGC 6522 belongs to a class of dynamically 
evolved clusters, including HP 1, NGC 6540, and NGC 6558, 
that are unique to the inner spheroid. The 
current data indicate that the $K$ LF of giants in NGC 6522 has a power-law 
exponent that is consistent with other core-collapsed inner spheroid clusters.

	$(K, J-K)$ CMDs of bright giants in NGC 6642 have been presented by 
Davidge (2000a) and Minniti et al. (1995). The latter found a relatively 
shallow giant branch, suggesting that NGC 6642 is moderately metal-rich, while 
the former found that [M/H] = --1.9. Minniti (1995b) concluded that [Fe/H] $= 
-1.4 \pm 0.2$ based on the strengths of Mg and Fe indices in the visible 
spectra of cluster giants, and this is consistent with the metallicity measured 
from the slope of the giant branch in the current work. Thus, NGC 6642 is a 
moderately metal-poor cluster.

\vspace{0.3cm}
\subsection{Globular Clusters and the Formation of the Inner Galaxy}

	Stars with [Fe/H] $<$ --1 account for only a small fraction of 
the total field population in BW (e.g. McWilliam \& Rich 1994). However, 
the June 1999 compilation of Harris (1996) indicates that the ratio of 
clusters within 3 kpc of the GC with [Fe/H] $>$ --1 to those with [Fe/H] $<$ 
--1 is 21/22 = 0.95. Thus, when measured with respect to field stars of 
similar metallicity, the frequency of metal-poor clusters in the inner Galaxy 
is much higher than that of metal-rich clusters.

	Various models (e.g. Fall \& Rees 1985; Harris \& Pudritz 1994, and 
McLaughlin \& Pudritz 1996) argue that globular cluster formation requires 
conditions that occur during the initial phases of protogalactic collapse. 
Another prediction is that globular clusters form before the field population 
(e.g. Fall \& Rees 1985), and this can explain why globular clusters in 
external galaxies tend to be more metal-poor than the underlying 
starlight (Forbes et al. 1996; Da Costa \& Mould 
1988). These formation models, when considered with observation 
evidence, suggest that globular clusters are not good 
{\it direct} tracers of the overall star-forming histories of galaxies (van den 
Bergh 1998), and the difference in the relative frequencies of metal-poor and 
metal-rich clusters supports this notion. However, the relative 
frequencies of metal-poor and metal-rich clusters may still
provide {\it indirect} clues into the early evolution of the inner spheroid. 

	Harris, Harris, \& McLaughlin (1998) investigated the specific globular 
cluster frequencies of brightest cluster galaxies, and suggested that the 
suppression of star formation after an initially large burst of activity, 
during which time globular clusters formed and star-forming material was 
ejected by stellar and supernovae-driven winds, could be 
responsible for the high specific cluster frequencies among these systems. 
When considered in the context of the Harris et al. (1998) 
scenario, the relative frequencies of metal-poor and metal-rich clusters 
in the inner Galaxy could be interpreted as being due to an early large 
spike in the star formation rate, at which time the metal-poor clusters formed. 
The formation of metal-poor field stars was stopped early-on 
by the ejection of gas, but star-forming activity was subsequently 
rejuvenated when chemically-enriched gas fell back into the central 
Galaxy, at which time the metal-rich clusters and field stars formed.

	Wyse \& Gilmore (1992) discuss the possible sources of gas from which 
the bulge formed, and argue that this material could not have come from the 
halo. If the metal-poor globular clusters in the inner spheroid formed during 
an initial spike in the star-formation rate, as suggested above, then 
some bulge stars may have formed from the metal-enriched gas that was 
ejected by SNII following this event, and so should have 
non-solar abundances of elements such as oxygen (see also Wyse \& Gilmore 
1992). McWilliam \& Rich (1994) find that at least some chemical 
elements in bulge giants are enriched with respect to solar.

	The scenario described above predicts that metal-poor clusters are 
older than metal-rich clusters, and the existence of such an 
age-metallicity relation remains a matter of controversy. While some studies 
find at least marginal evidence for such a relation (e.g. Chaboyer, Demarque, 
\& Sarajedini 1996; Richer et al. 1996; Salaris \& Weiss 1998; Rosenberg et al. 
1999), others conclude that the metal-rich and metal-poor clusters are 
coeval (e.g. Ortolani et al. 1995; Buonanno et al. 1998). The various sources 
of uncertainty that frustrate efforts to compare the ages of clusters spanning 
a broad range of metallicities are discussed by Rosenberg et al. (1999).

	The chemical composition of cluster stars provides broad constraints on 
the time delay between the formation of the metal-poor and metal-rich clusters, 
and the existing data, while not conclusive, suggest that any age difference 
may prove difficult to measure. If the metal-poor and metal-rich clusters 
have similar $\alpha$ element abundances, as suggested by Carney (1996), then 
the clusters formed from material that was enriched on a timescale that is 
shorter than the onset of SN I ($\leq 1$ Gyr), and age differences of this 
size would be challenging to detect. However, if the metal-rich clusters show a 
trend of decreasing $\alpha$ enrichment towards higher metallicities, as is the 
case among field stars and may be the case among globular clusters (Ferraro et 
al. 1999), then the cluster system formed from material that was 
enriched over a longer time span. 

	The mass of gas that was initially present 
in the inner Galaxy can be estimated knowing (1) the total mass of globular 
clusters, and (2) the globular cluster formation 
efficiency, $\epsilon$, which is the ratio of mass in globular clusters to that 
in stars and gas (McLaughlin 1999). As noted by McLaughlin (1999), $\epsilon$ 
is a more physical measure of cluster statistics than the traditional specific 
frequency, S$_N$, as $\epsilon$ accounts for sources of mass that are not 
accounted for in S$_N$, such as gas in extended halos. 
Harris et al. (1998) argue that $\epsilon$ is a constant that holds over a 
range of different environments.

	There are 22 metal-poor and 21 metal-rich globular clusters with 
R$_{GC} \leq 3$ kpc (Harris 1996) and, if each has a mass of $2 \times 10^5$ 
M$_{\odot}$ (McLaughlin 1999), then the total mass of these objects is:

\hspace*{5.0cm} $43 \times 2 \times 10^5 = 8.6 \times 10^6$ M$_{\odot}$

	This is only a lower limit to the total mass of the early cluster 
system, since (1) globular clusters in the inner Galaxy are dissrupted by 
dynamical interactions, and (2) incompleteness in the present-day cluster 
population has not been accounted for, although the number of missed clusters 
is likely not large, and will not be considered further. 
Simulations conducted by Vesperini (1997) suggest that only a 
third of the current cluster population has survived in the central 3 kpc. 
Therefore, the total initial mass of clusters is:

\hspace*{5.0cm} $3 \times 8.6 \times 10^6 = 2.6 \times 10^7$ M$_{\odot}$

	McLaughlin (1999) computed $\epsilon$ for three massive 
early-type galaxies, and found a mean value $\overline{\epsilon} = 0.0027$
with only modest galaxy-to-galaxy scatter. If this value of 
$\epsilon$ is also assumed to hold for the inner spheroid cluster system, 
then the globular clusters formed from an initial supply of gas that had a mass:

\hspace*{5.0cm} $\frac{2.6 \times 10^7}{2.7 \times 10^{-3}} = 10^{10}$ M$_{\odot}$

	The mass of the bulge within 3 kpc is roughly $10^{10}$ 
M$_{\odot}$ (Blanco \& Terndrup 1989), and this is in remarkable 
agreement with the gas mass computed above. It thus appears that 
during the earliest episodes of star formation in the inner Galaxy 
there was a large reservoir of material from which a 
significant fraction of the current bulge population could have been 
assembled. Of course, star formation is an inefficient process, and only some 
fraction of the available gas will be turned into stars. Adopting the 
relatively high efficiency of 10\% that is required to produce bound clusters 
(Elmegreen \& Clemens 1985), then only $10^9$ M$_\odot$ of stars would 
be formed during any large-scale star-forming episode in the inner spheroid; 
however, the cumulative efficiency due to subsequent star-forming episodes will 
be much higher if gas is re-cycled. These admittedly simple 
calculations indicate that it may not be necessary to invoke a separate 
pre-enriched body of gas to form a large fraction of the bulge; rather, the 
early enrichment of this material may have occured {\it in situ}.

\vspace{0.3cm}
	Sincere thanks are extended to Sidney van den Bergh for commenting 
on an earlier version of this paper, and to Peter Stetson for discussions 
concerning the ages of globular clusters. An anonymous referee also provided 
comments that greatly improved the paper.

\clearpage

\begin{table*}
\begin{center}
\begin{tabular}{lcccc}
\tableline\tableline
Cluster & [Fe/H] & E(B--V) & R$_{GC}$ & Core-Collapsed? \\
 & & & (kpc) & (Yes/No) \\
\tableline
NGC 6273 & --1.68 & 0.37 & 1.6 & N \\
NGC 6287 & --2.05 & 0.60 & 1.7 & N \\
NGC 6293 & --1.92 & 0.41 & 1.4 & Y \\
NGC 6333 & --1.72 & 0.38 & 1.7 & N \\
NGC 6355 & --1.50 & 0.75 & 1.0 & Y \\
NGC 6401 & --1.12 & 0.85 & 0.8 & N \\
NGC 6522 & --1.44 & 0.48 & 0.6 & Y \\
NGC 6642 & --1.35 & 0.41 & 1.6 & Y? \\
 & & & & \\
NGC 288 & --1.24 & 0.03 & 11.6 & N \\
M13 & --1.54 & 0.02 & 8.7 & N \\
M92 & --2.29 & 0.02 & 9.6 & N \\
\tableline
\end{tabular}
\end{center}
\caption{CLUSTER PROPERTIES}
\end{table*}

\clearpage

\begin{table*}
\begin{center}
\begin{tabular}{lcc}
\tableline\tableline
Cluster & $\frac{\Delta(J-K)}{\Delta(K)}$ & [M/H]$_{Ferraro et al.}$ \\
\tableline
NGC 6273 & $-0.069 \pm 0.002$ & $-1.4 \pm 0.2$ \\
NGC 6287 & $-0.031 \pm 0.004$ & $-2.3 \pm 0.2$ \\
NGC 6293 & $-0.033 \pm 0.004$ & $-2.2 \pm 0.2$ \\
NGC 6333 & $-0.037 \pm 0.003$ & $-2.1 \pm 0.2$ \\
NGC 6355 & $-0.041 \pm 0.005$ & $-2.0 \pm 0.2$ \\
NGC 6401 & $-0.065 \pm 0.003$ & $-1.5 \pm 0.2$ \\
NGC 6522 & $-0.063 \pm 0.004$ & $-1.5 \pm 0.2$ \\
NGC 6642 & $-0.055 \pm 0.005$ & $-1.7 \pm 0.2$ \\
 & & \\
NGC 288 & $-0.054 \pm 0.006$ & $-$ \\
M13 & $-0.049 \pm 0.004$ & $-1.8 \pm 0.2$ \\
M92 & $-0.035 \pm 0.006$ & $-2.2 \pm 0.2$ \\
\tableline
\end{tabular}
\end{center}
\caption{GIANT BRANCH SLOPES AND METALLICITIES}
\end{table*}

\clearpage

\begin{table*}
\begin{center}
\begin{tabular}{ccc}
\tableline\tableline
Cluster & $x$ & $x_{Davidge 2000a}$ \\
\tableline
NGC 6273 & $0.34 \pm 0.03$ & $0.23 \pm 0.02$ \\
NGC 6287 & $0.22 \pm 0.03$ & $0.38 \pm 0.02$ \\
NGC 6293 & $0.28 \pm 0.02$ & $0.25 \pm 0.07$ \\
NGC 6333 & $0.40 \pm 0.02$ & $0.20 \pm 0.08$ \\
NGC 6355 & $0.26 \pm 0.02$ & $0.19 \pm 0.07$ \\
NGC 6401 & $0.41 \pm 0.03$ & -- \\
NGC 6522 & $0.27 \pm 0.04$ & $0.31 \pm 0.07$ \\
NGC 6642 & $0.31 \pm 0.04$ & $0.17 \pm 0.03$ \\
 & & \\
NGC 288 & $0.24 \pm 0.06$ & $0.22 \pm 0.03$ \\
M13 & $0.34 \pm 0.05$ & -- \\
M92 & $0.29 \pm 0.02$ & -- \\
\tableline
\end{tabular}
\end{center}
\caption{LUMINOSITY FUNCTION EXPONENTS}
\end{table*}

\clearpage

\begin{table*}
\begin{center}
\begin{tabular}{cccc}
\tableline\tableline
Cluster & $\mu_0^{RGB-bump}$ & $\mu_0^{RGB-tip}$ & $\mu_0^{HB}$ \\
\tableline
NGC 6273 & 14.8 & 14.1 & 14.5 \\
NGC 6401 & 14.6 & -- & -- \\
NGC 6522 & 15.1 & 14.4 & 14.1 \\
NGC 6642 & 15.2 & 15.6 & 14.3 \\
\tableline
\end{tabular}
\end{center}
\caption{CLUSTER DISTANCES}
\end{table*}

\clearpage

\begin{center}
FIGURE CAPTIONS
\end{center}

\figcaption
[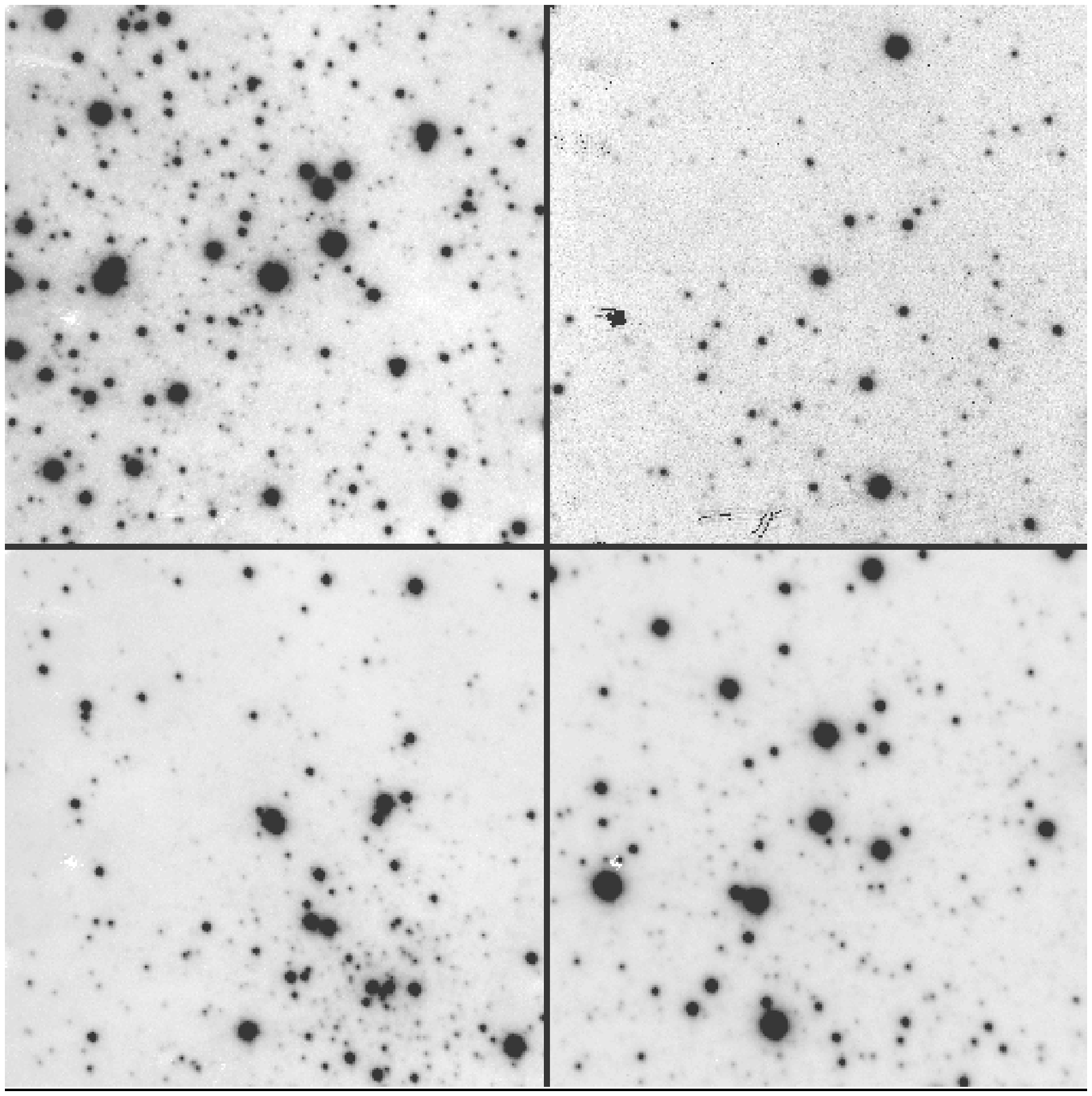]
{The final deep $K$ images of NGC 6273 (upper left hand corner), NGC 6287 
(upper right hand corner), NGC 6293 (lower left hand corner), and NGC 
6333 (lower right hand corner). Each image is $34 \times 34$ arcsec, with 
North at the top, and East to the left. 
The AO reference stars are located near the field centers.}

\figcaption
[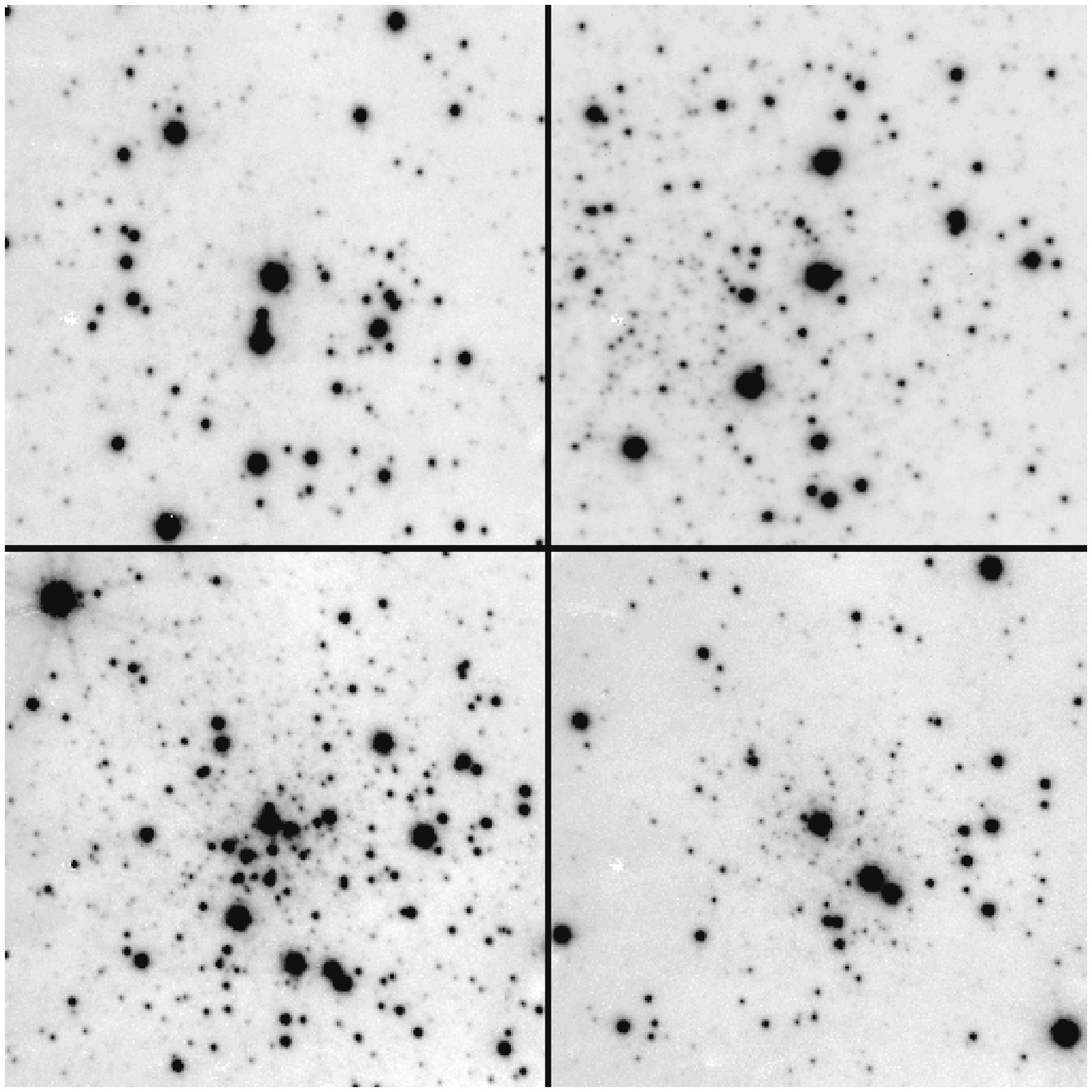]
{The final deep $K$ images of NGC 6355 (upper left hand corner), NGC 6401 
(upper right hand corner), NGC 6522 (lower left hand corner), and NGC 
6642 (lower right hand corner). Each image is $34 \times 34$ arcsec, with 
North at the top, and East to the left. 
The AO reference stars are located near the field centers.}

\figcaption
[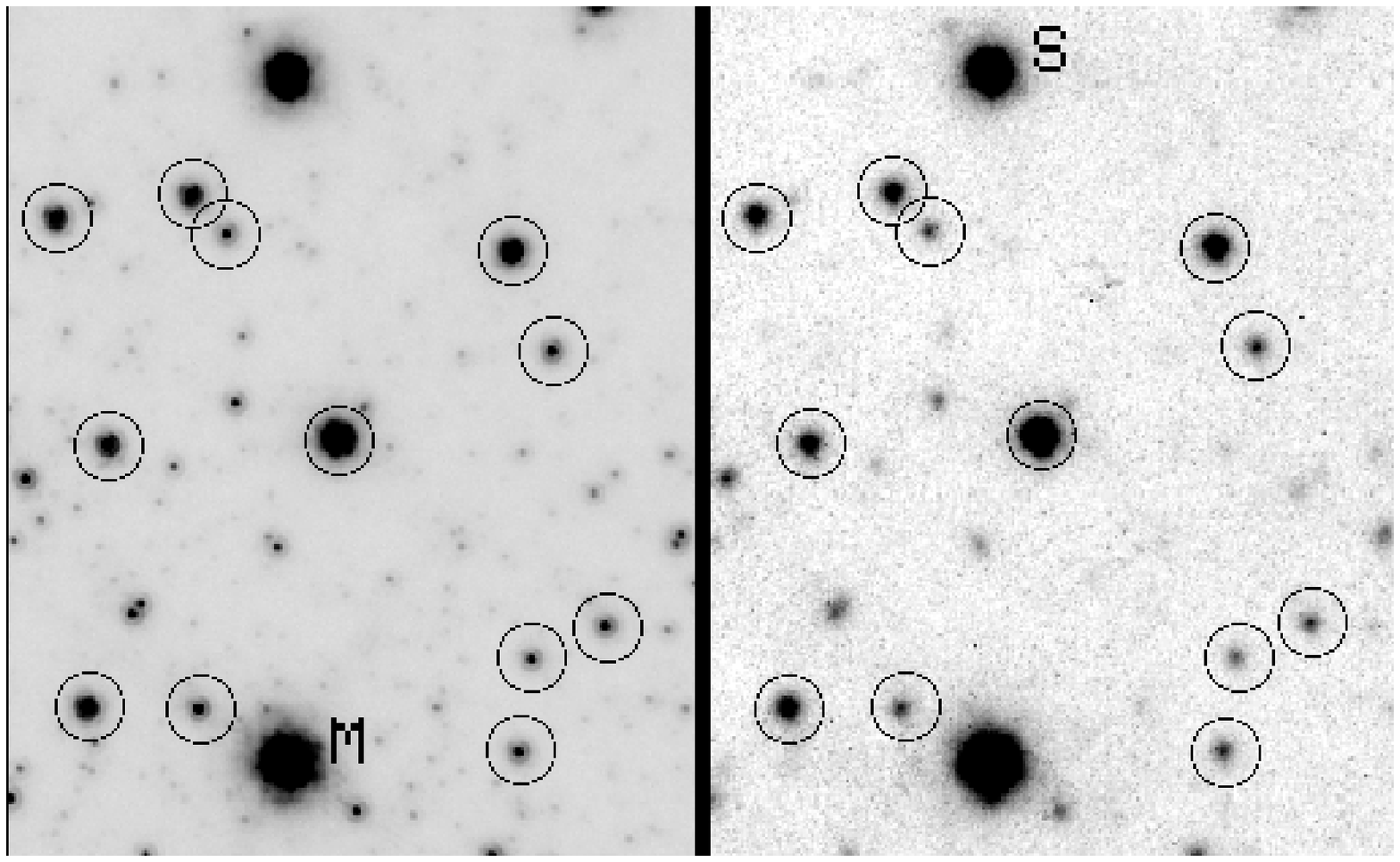]
{$K$ images of a field near the center of NGC 6287, obtained 
during March 1998 (left hand panel) and September 1998 (right hand panel). 
The guide star for the March 1998 observations is marked with an `M', while 
the guide star for the September 1998 observations is marked with an `S'. The 
stars used to investigate the effects of anisoplanicity on the photometric 
measurements are indicated with circles.}

\figcaption
[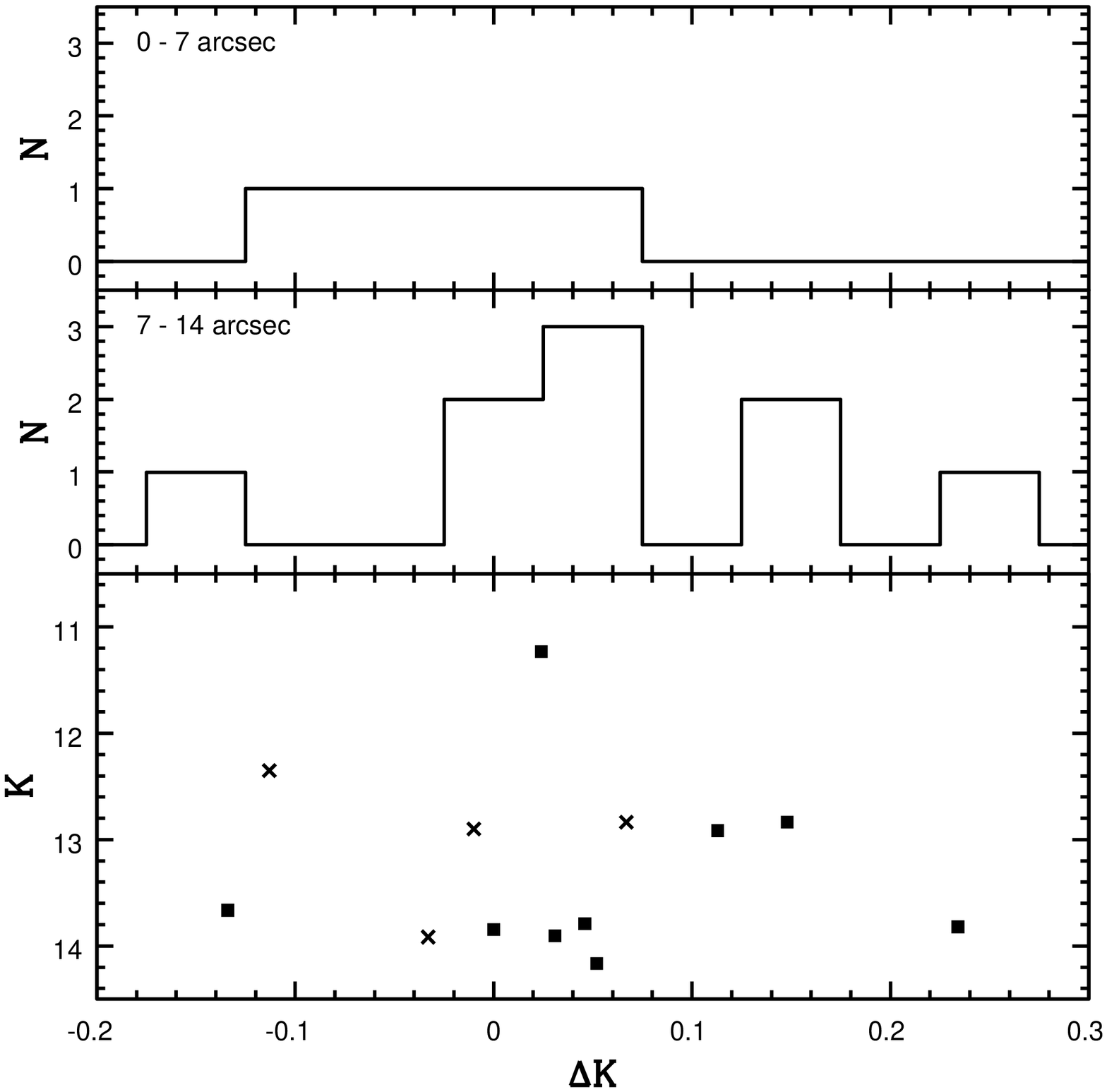]
{The top two panels show the histogram distributions of the difference in $K$ 
brightness measured from the March 1998 and September 1998 datasets for 
the stars marked in Figure 3. The angular intervals refer to 
distances from star `S' in Figure 3. The means of the two 
distributions differ at only the $1.6\sigma$ significance level. The lower 
panel shows the relation between $K$ and $\Delta K$; crosses are stars 
within 7 arcsec of star `S', while the solid squares are stars 
located between 7 and 14 arcsec of this star. Note that the extreme values 
in $\Delta K$ occur for stars that are near the faint limit of the 
September 1998 dataset.}

\figcaption
[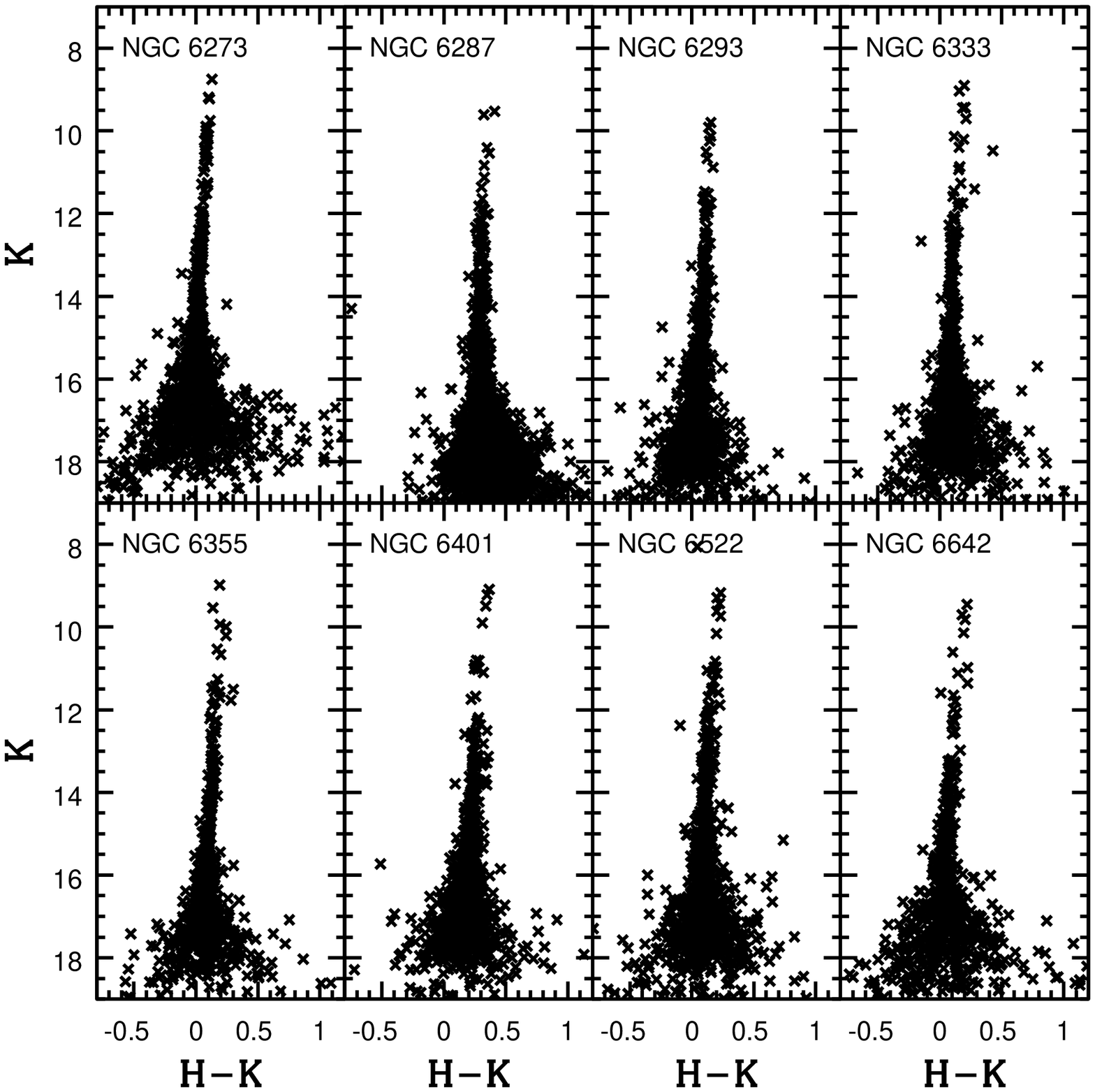]
{The $(K, H-K)$ CMDs of the inner spheroid clusters. Note the 
modest scatter at intermediate brightnesses; the standard deviation in $H-K$ 
for stars between $K = 12$ and 14 is typically $\pm 0.02$ to $\pm 0.03$ mag, 
which is consistent with that predicted from artificial star experiments. 
The NGC 6287 data are from Davidge \& Courteau (1999).}

\figcaption
[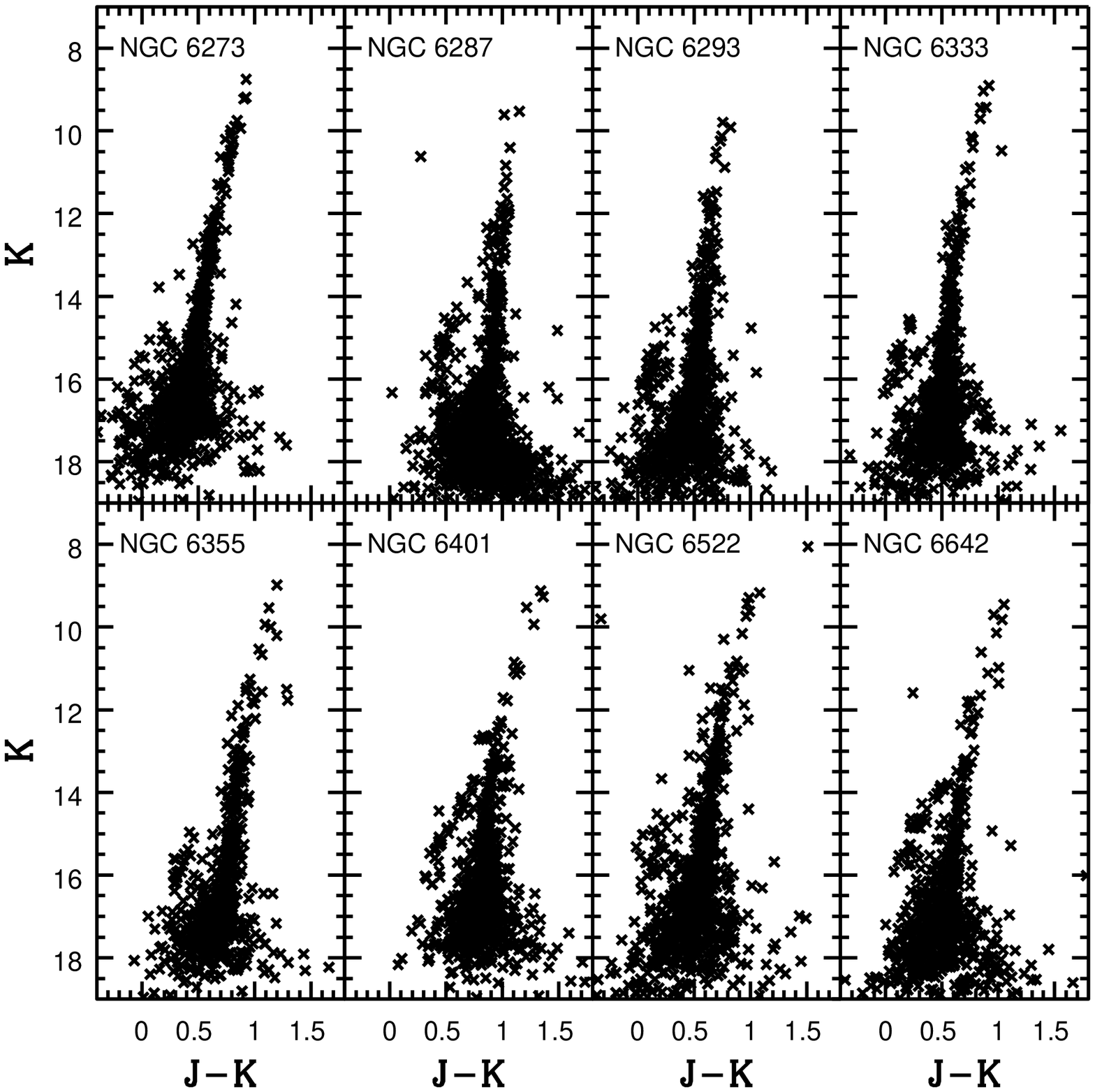]
{The $(K, J-K)$ CMDs of the inner spheroid clusters. The CMDs 
terminate near the MSTO, and the HB can be 
seen to the left of the RGB between K = 13 and 16. 
The NGC 6287 data are from Davidge \& Courteau (1999).}

\figcaption
[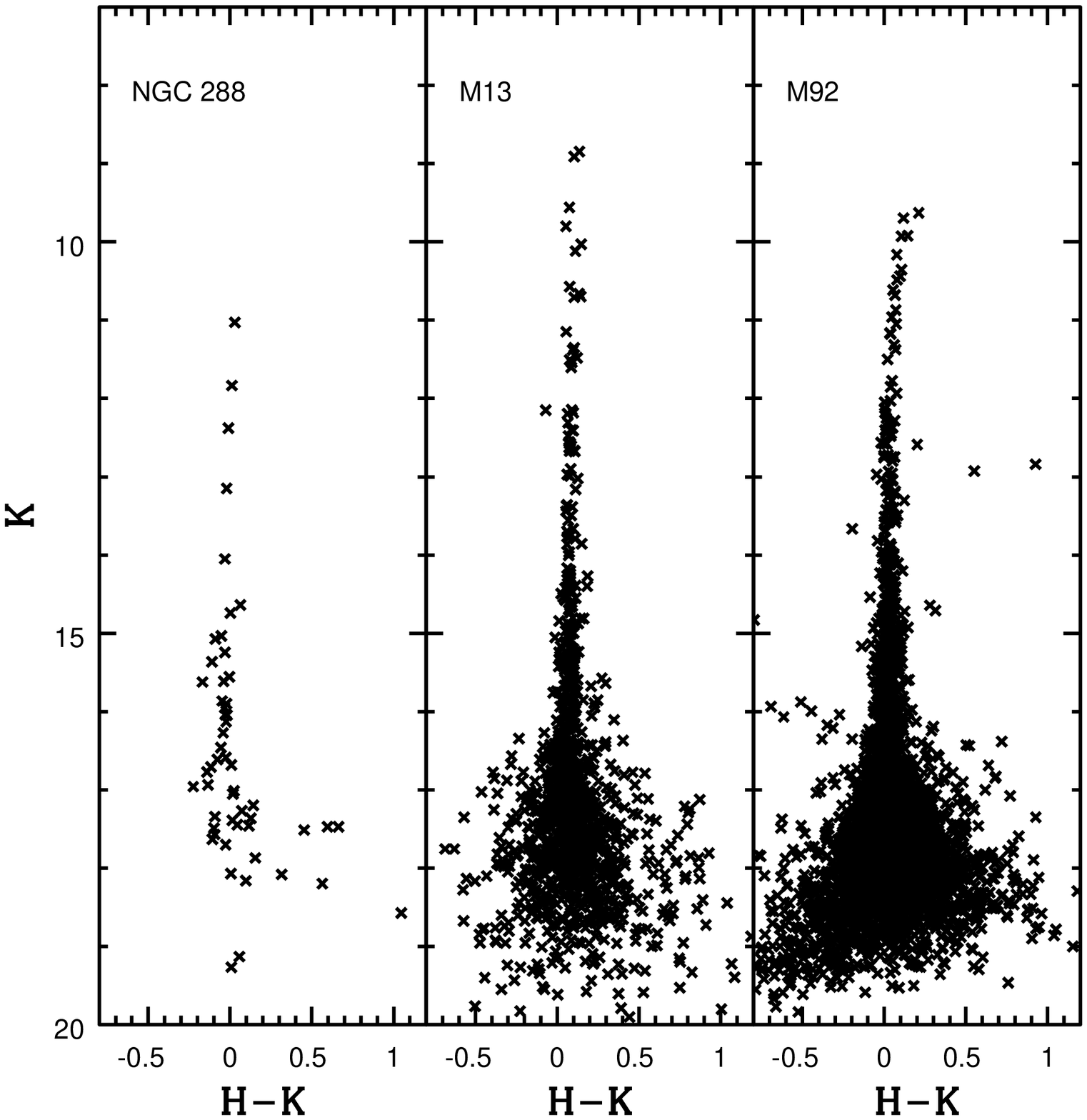]
{The $(K, H-K)$ CMDs of the outer halo clusters. The 
M13 and M92 data are from Davidge \& Courteau (1999).}

\figcaption
[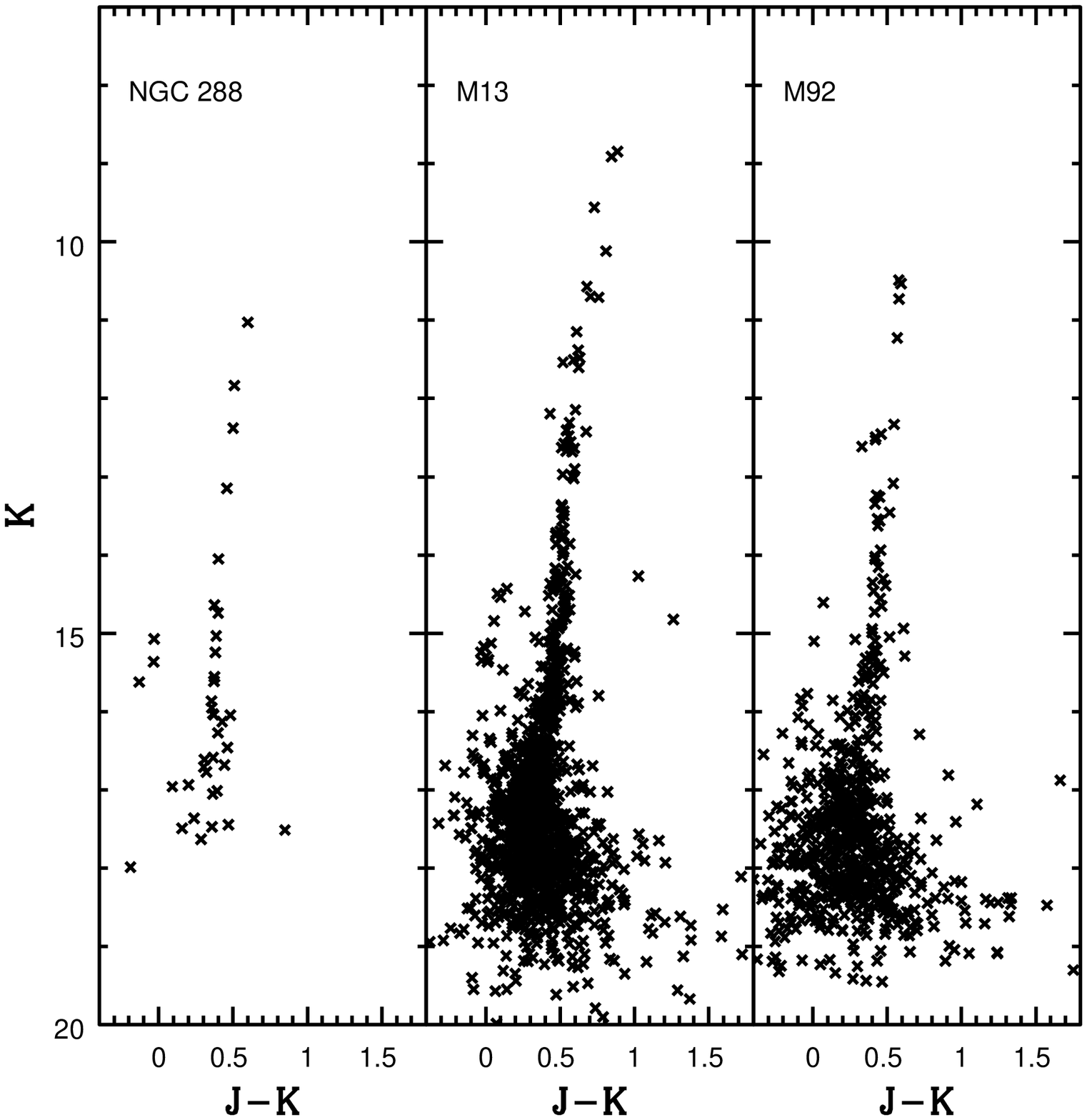]
{The $(K, J-K)$ CMDs of the outer halo clusters.
The M13 and M92 data are from Davidge \& Courteau (1999).}

\figcaption
[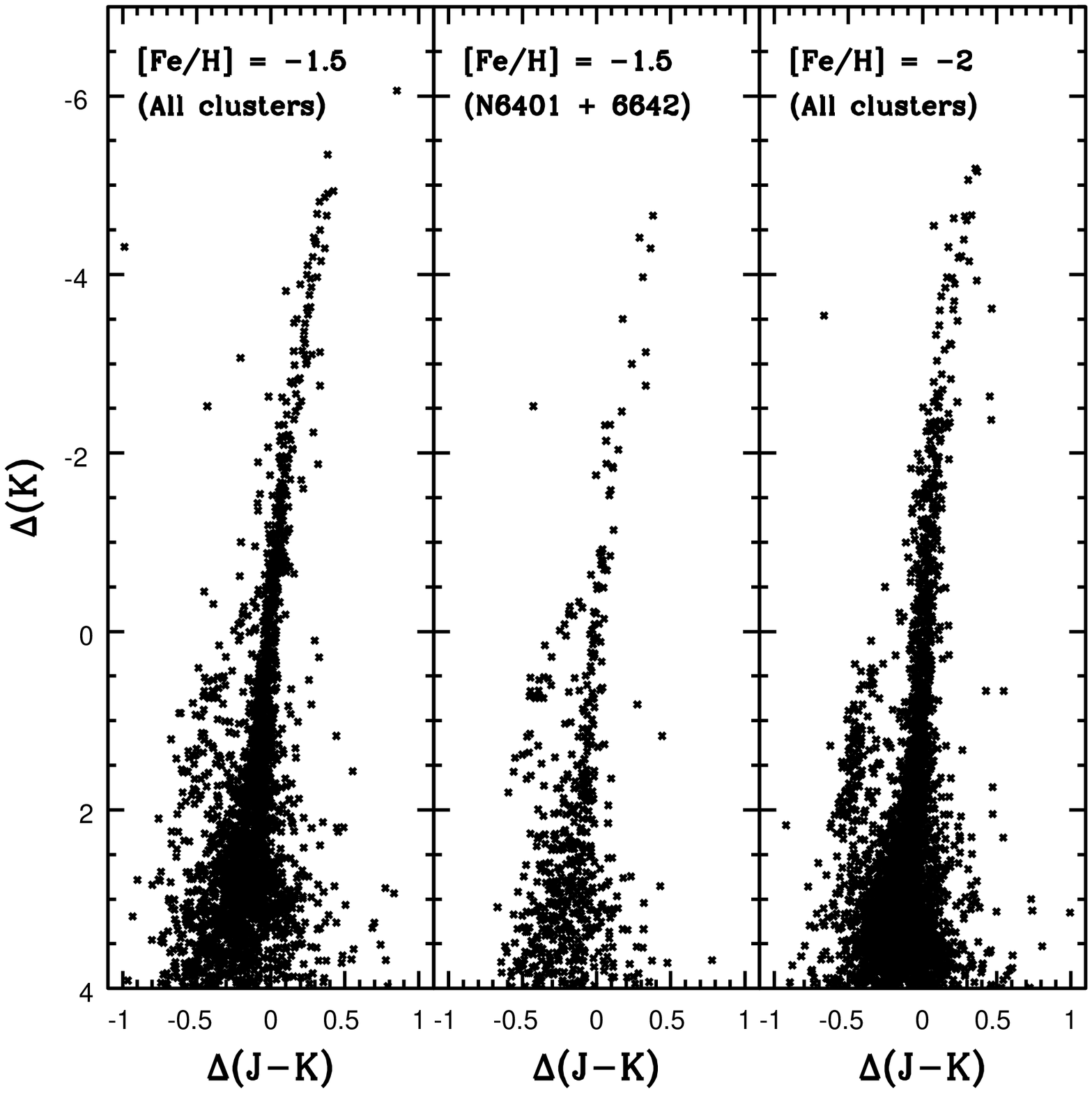]
{Composite $(K, J-K)$ CMDs of the inner spheroid clusters grouped according to 
metallicity. The left hand panel shows the CMDs of NGC 6273, NGC 6401, NGC 
6522, and NGC 6642, while the central panel shows the CMDs of NGC 6401 and NGC 
6642, which have the best-defined HB sequences in the moderately metal-poor 
group. The right hand panel shows the CMDs of NGC 6287, NGC 6293, NGC 6333, and 
NGC 6355. Cluster-to-cluster differences in reddening have been removed 
using the procedure described in the text. 
$\Delta$(K) and $\Delta$(J-K) are measured from the 
shifted giant branches at K = 14. There are clear differences between 
the HBs of the [Fe/H] $= -1.5$ and [Fe/H] $= -2$ groups.}

\figcaption
[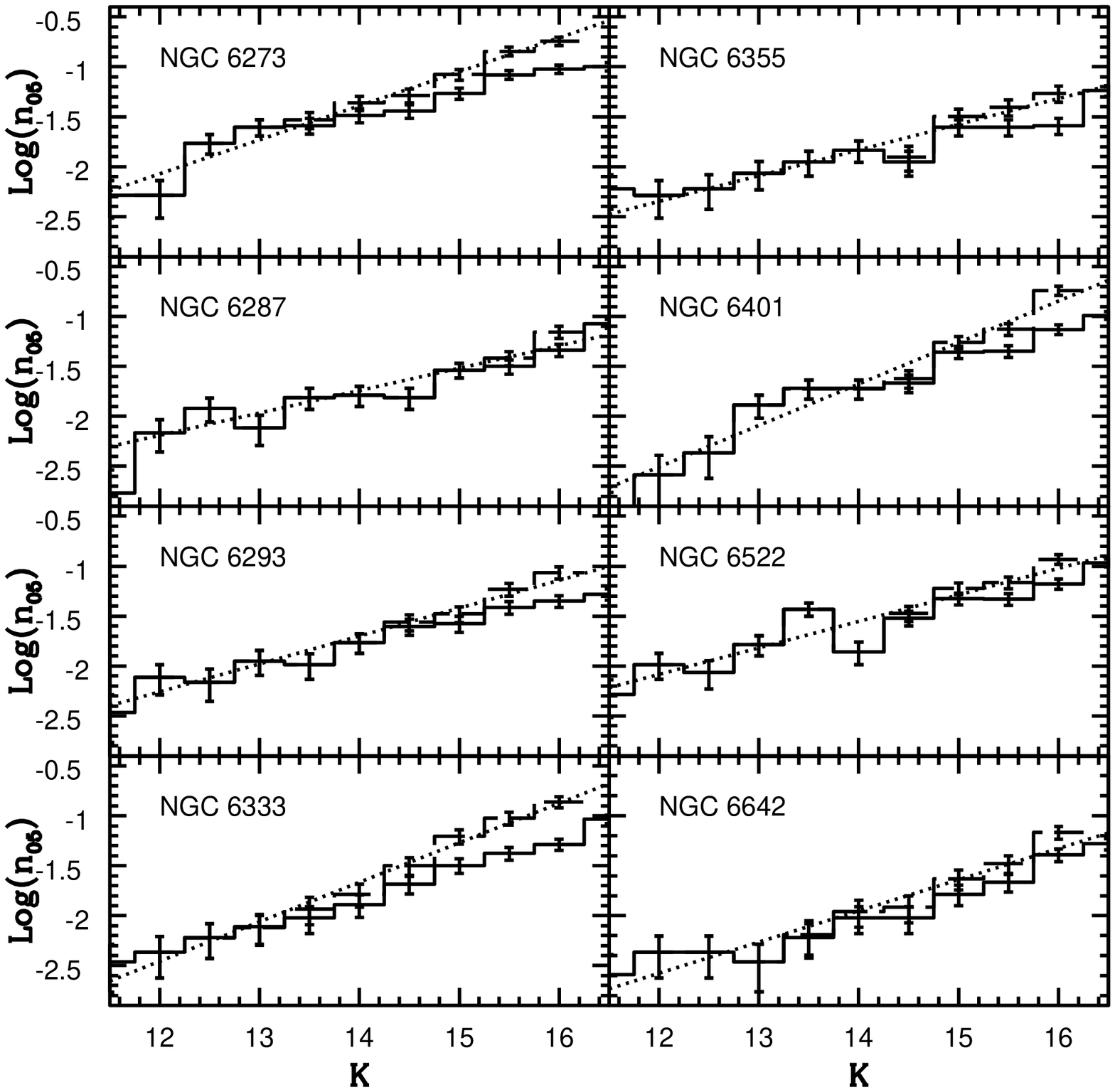]
{The $K$ LFs of giant branch stars in inner spheroid clusters. The LFs were 
constructed from the CMDs, and HB stars have been removed. 
The solid lines are the raw LFs, while the dashed lines are the LFs 
corrected for incompleteness. n$_{05}$ is the number of stars per 0.5 magnitude 
interval per square arcsec. The dotted lines are power-laws, the exponents for 
which are listed in the second column of Table 3, that were fit to the 
completeness-corrected LFs between $K = 12$ and $K = 16$.}

\figcaption
[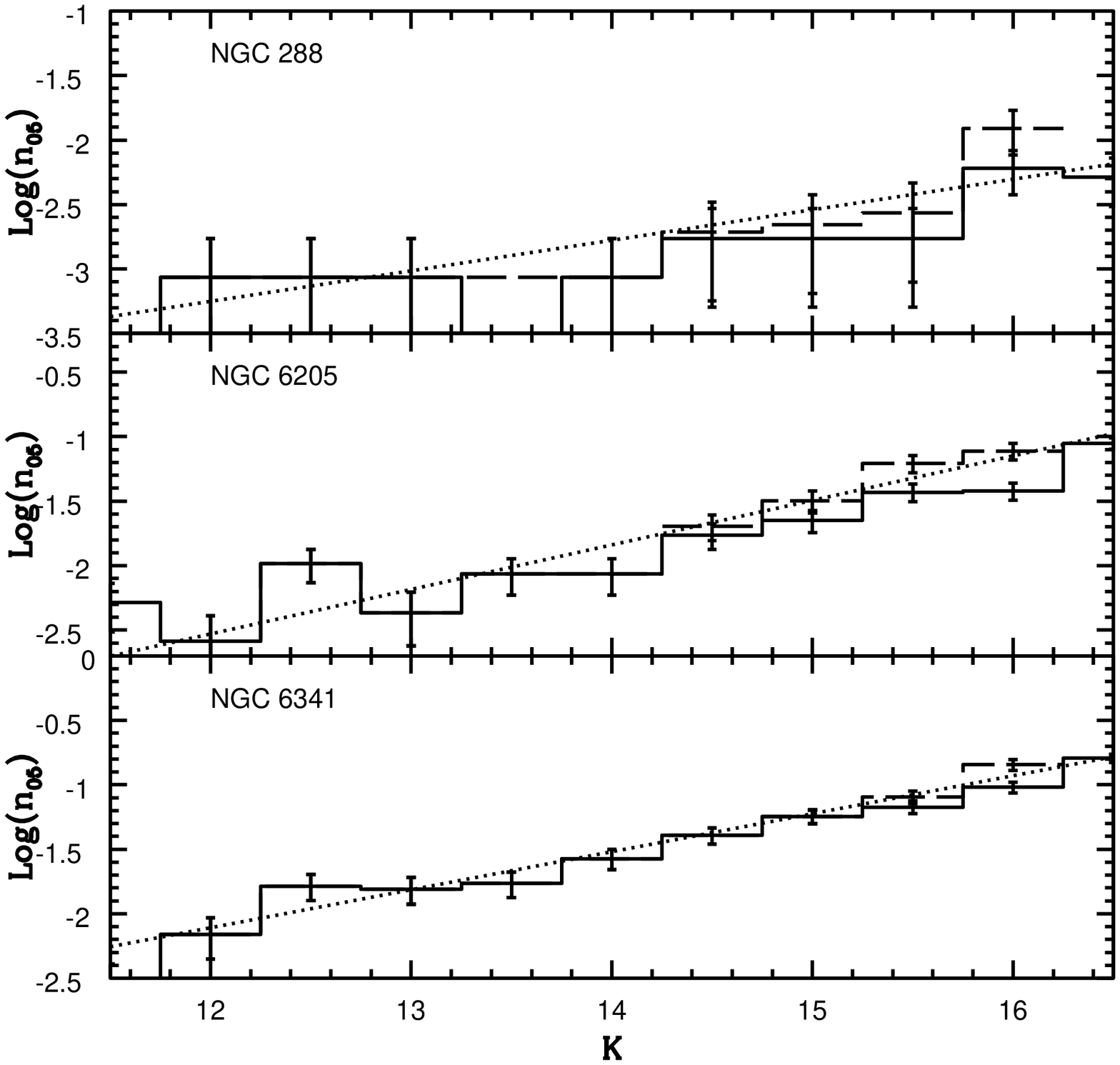]
{Same as Figure 10, but for the outer halo clusters.}

\figcaption
[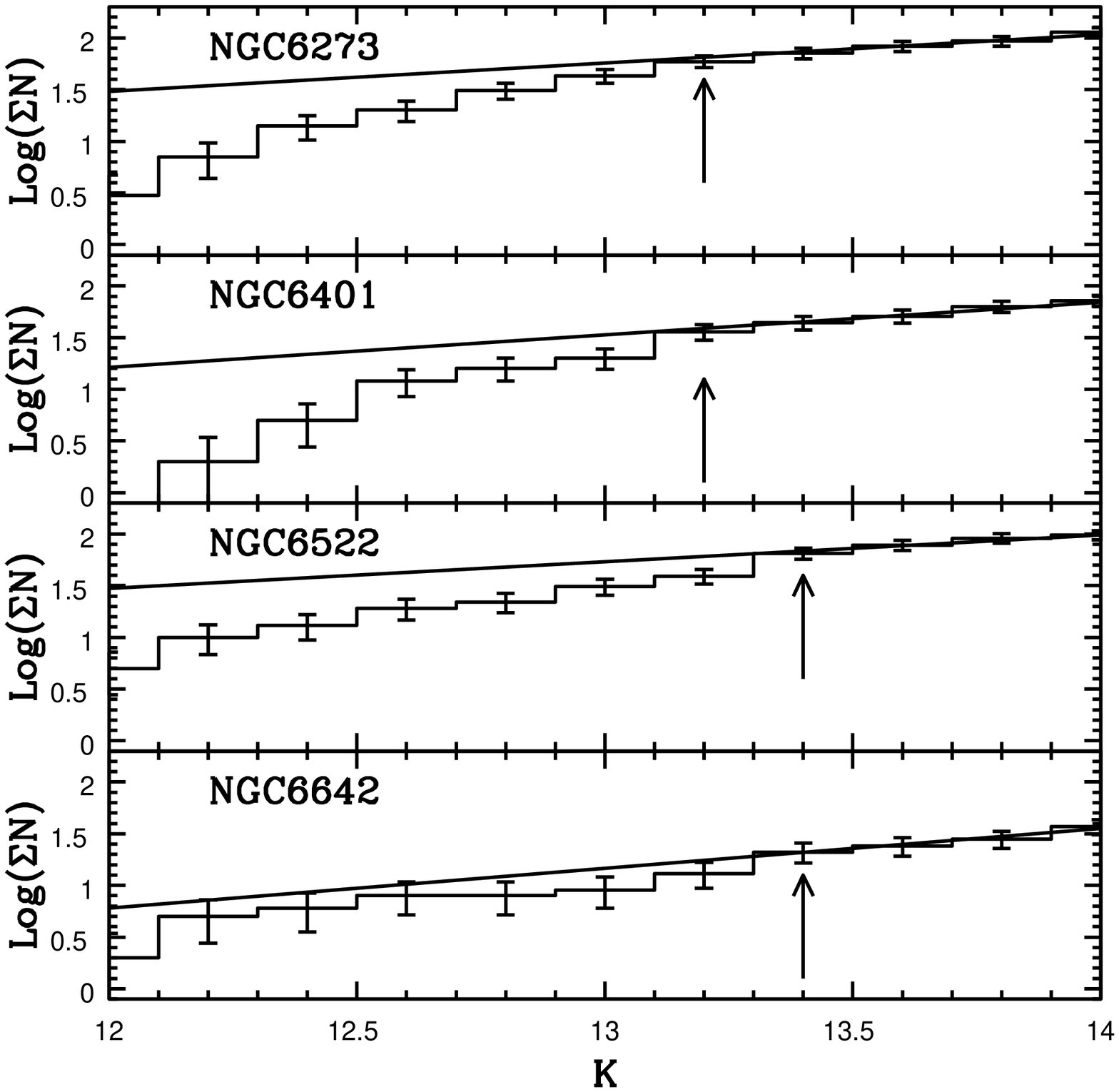]
{The integrated $K$ LFs of bright giant branch stars in NGC 6273, NGC 6401, NGC 
6522, and NGC 6642, where secure RGB-bump detections have been made. 
$\Sigma$N$_{02}$ is the cumulative number of stars per 0.2 magnitude interval, 
and the error bars show the uncertainties introduced by counting statistics. 
The arrows mark the location of the RGB-bump, which is defined as the 
point after which the star counts depart from the relation predicted by stars 
between $K = 13.5$ and 15, which is shown as a solid line.} 

\figcaption
[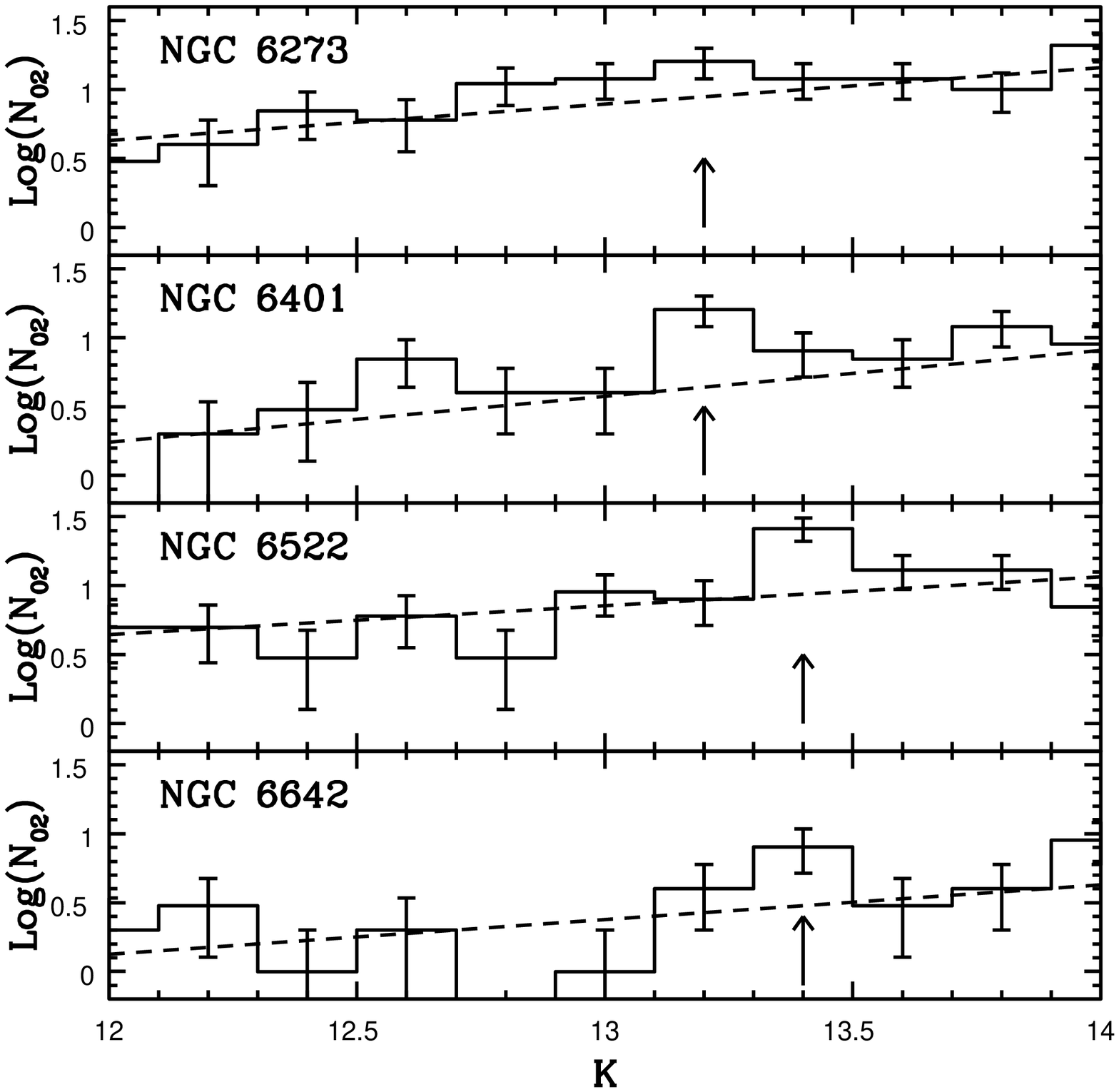]
{The differential $K$ LFs of bright giant branch stars in those clusters for 
which the RGB-bump has been identified. N$_{02}$ is the number of stars per 0.2 
magnitude interval, and the error bars show the uncertainties introduced by 
counting statistics. The dashed line shows a least squares fit to the giant 
branch LF between $K = 12$ and 16 for each cluster, and the arrows mark the 
location of the RGB-bump predicted from the integrated LFs in Figure 12. In all 
four clusters the star counts at the expected RGB-bump brightness exceed, at 
the $2-\sigma$ or higher significance level, those expected from the least 
squares fit, thus confirming the detection of this feature.} 

\end{document}